# Mapping Evolution of Molecules Across Biochemistry with Assembly Theory

Sebastian Pagel, Abhishek Sharma, and Leroy Cronin*

School of Chemistry, The University of Glasgow, University Avenue, Glasgow G12 8QQ, UK

*Corresponding author email: Lee.Cronin@glasgow.ac.uk

**Abstract**

Evolution is often understood through genetic mutations driving changes in an organism's fitness, but there is potential to extend this understanding beyond the genetic code. We propose that natural products—complex molecules central to Earth's biochemistry—can be used to uncover evolutionary mechanisms beyond genes. By applying Assembly Theory (AT), which views selection as a process not limited to biological systems, we can map and measure evolutionary forces in these molecules. AT enables the exploration of the assembly space of natural products, demonstrating how the principles of the selfish gene apply to these complex chemical structures, selecting vastly improbable and complex molecules from a vast space of possibilities. By comparing natural products with a broader molecular database, we can assess the degree of evolutionary contingency, providing insight into how molecular novelty emerges and persists. This approach not only quantifies evolutionary selection at the molecular level but also offers a new avenue for drug discovery by exploring the molecular assembly spaces of natural products. Our method provides a fresh perspective on measuring the evolutionary processes both, shaping and being read out, by the molecular imprint of selection.

**Introduction**

The theory of evolution allows us to elucidate the processes by which distinct populations of species arise, finely tailored to occupy specific biological niches within a given biosphere *(1)*. If there are sufficient resources, the species survives and reproduces until superseded by a better-adapted species or until the resources diminish and this represents the process of natural selection. While evolutionary theory describes the change of organisms on a biological level, the underlying mechanistic dynamics of evolutionary and pre-evolutionary processes at the molecular scale remain to be defined. The key physical phenomenon ubiquitous at all scales from molecules to complex evolutionary architectures such as cells is *selection*. Previously, various approaches have been utilised to describe Darwinian evolution such as the quasi-species model based on a physical chemistry framework *(2)*, evolutionary game theory *(3, 4)* and minimal cellular models such as the Chemoton *(5)*. Most of these models probe evolutionary dynamics by defining key processes at the molecular scale such as replication-mutation leading to natural selection by processes such as cooperation, competition etc. However, these models cannot give full mechanistic insight into the process of selection within the vast chemical space starting at the fundamental molecular scale leading to the emergence *(6)* of various biochemical processes including autocatalytic sets *(7)*, replicators *(8)*, and metabolic pathways *(9)*. Quantifying the degree of selection required at the molecular scale to create biologically relevant molecules *(10)* is a complex and open problem.

Recently, Assembly Theory (AT) was utilised to distinguish biological from non-biological samples, showing that complex molecules if observed in high abundance can act as biosignatures for life on Earth, and this represents a generalised approach to search for life *(11–13)*. The observation of complex molecules is the outcome of evolutionary processes that occur within all biological life as we know it. As an example, within the Earth's biosphere, complex molecules can either be functional as is NADP+, which is constantly being synthesised, transformed and regenerated *(14)*, or as secondary metabolites that are not directly involved in the survival of an organism but are often characteristic of a specific genus, species or strain *(15)*. Well-known examples of plant secondary metabolites can be categorised as terpenoids, phenolic compounds, sulphur-containing compounds, and alkaloids *(16)* for example papaverine *(17)* or paclitaxel *(18)*. The production of these complex molecules appears to be driven by the evolutionary dynamics within the ecosystem, whereby these secondary metabolites can have a causal influence on other living systems *(19)*. While primary metabolites such as amino acids, organic acids etc. *(20)* function as all kinds of building blocks in cellular processes, secondary metabolites *(15,*

*16)* interact with external agents and stresses *(21–24)* giving them selective dominance which could lead to higher chances of survival. Consequently, secondary metabolites accumulate in these organisms, and the environment they inhabit *(22)*. The combination of complexity and accumulation (high copy number in the framework of AT) makes secondary metabolites the premier biosignatures and markers of selection since they allow for an almost unambiguous proof of chemical causality only enforced by a strongly selective evolutionary machine, as observed on Earth *(11)*. This is because such complex molecules would not spontaneously arise, but their synthesis pathways have been selected within a population of organisms in each environment. Indeed, the production of natural products can be driven by natural selection, where they can provide an advantage to the organism, such as defending against predators, attracting pollinators, or competing with other species in a molecular arms race.

The quest to find a generalised definition of what life would look like on macroscopic and molecular scales remains an open research question without much consensus. As a potential solution, AT was further extended to quantify and generalise *selection* in physical systems beyond natural selection as defined in biological systems *(13)*. AT introduces the concept of an object (such as a molecule) as an entity which is finite, distinguishable, persists in time, and is breakable such that the set of constraints to construct it are quantifiable. Thus, the complexity of an object is quantified by defining an intrinsic measure called the assembly index ($a$) which is the shortest number of steps to construct an object in the absence of any physical constraints. For an ensemble of objects, we have defined an integrated quantity Assembly ($A$) of the ensemble which signifies the total amount of selection necessary to produce a set of observed objects, quantified using equation (1):

$$A = \sum_{i=1}^{N} e^{a_i} \left( \frac{n_i - 1}{N_T} \right) \quad (1)$$

where $a_i$ is the assembly index of $i^{th}$ object, $n_i$ is its copy number or concentration and $N_T$ the total number of objects in the ensemble *(13)*. The combination of assembly index and copy number quantifies within the large combinatorial space, how selection process leads to the discovery of new objects and among them the system's capacity to produce a high copy number of specific objects.

Herein, we present a detailed exploration of the application of Assembly Theory (AT) to the study of natural products, with the aim of quantifying the evolutionary processes that shape complex biochemical systems. Through our analysis, we demonstrate that the assembly spaces of natural products, as derived from the COCONUT database, represent highly selective outcomes of evolutionary processes. By comparing these assembly spaces with those of all known molecules, we quantify the degree of selection required to generate the observed molecular diversity on Earth. We also explore how a systematic reduction in the constraints controlling these assembly spaces can lead to the generation of novel molecular structures under alternate selection pressures, that is giving new constraints not limited by the previous ones. This approach is particularly significant for drug discovery, as it allows us to reconstruct and explore drug-like molecules that retain key features of natural products, which have evolved to interact with biological systems. By mimicking the fragment utilization and construction step distribution observed in natural products, we can generate new chemical spaces that are enriched in molecules with desirable drug-like properties, while also incorporating evolutionary advantages inherent to natural products. This work provides a robust framework for understanding the intersection of selection, evolution, and molecular complexity, offering new insights into the origins and potential future trajectories of biochemical diversity, particularly in the context of drug discovery.

To use AT to explore natural product space, we first need to define the shortest pathway to construct the object as the *assembly pathway* on which the assembly index (molecular assembly index in the case of molecules) can be quantified. This is important because, in the absence of the knowledge of the mechanistic insights through which the object has been created, the assembly pathway represents the informational constraints along the construction pathway required to build the object. This approach, using assembly pathways, allows all the observed objects to be detected in a similar way such that the bias that emerges is due to the selective processes can be estimated by quantifying the historical contingency associated to the construction processes. Thus, the assembly pathways over an ensemble of objects constitute the combinatorial assembly spaces, see Fig. 1 *(13)*.

Building on the concept of an assembly space, *selection* is defined in AT as a temporal process leading to the transition from undirected to directed exploration dynamics. We can consider that a physical process represented by forward dynamics, in the assembly space, selection can be observed in time by measuring the selectivity parameter alpha ($\alpha$) which is equal to one for an undirected process and drops below one for a directed process *(13)*. Here we quantify the amount of selection required to construct the assembly space of molecules representing biochemical processes occurring on planet Earth. Given the lack of temporal information of the evolution of biochemical processes, we postulate that it is possible to use the molecules in the natural products database as an ensemble of complex objects which represent the Assembly Observed. Assuming that the database is samples most of all the biochemical processes and has been observed in high enough abundance, we assume that exploration occurred over a long evolutionary timescale, and the Assembly Observed of natural products gives a good representation of the natural products produced by evolution which is captured in the space of the Assembly Contingent and represented as $A_{NP}$ – that is the space of possible natural products. The Assembly Possible represents all physically plausible molecules and expands exponentially, and hence is computationally intractable. Instead for simplicity, we consider all the molecules found in the PubChem database to represent physically plausible molecules as represented as $A_M$ such that the space of natural products is a subset of the space of all physically possible molecules. The degree of selection is quantified by comparing features of assembly spaces of all possible molecules and natural products, see Fig. 1. We then extend the methodology to explore new chemical spaces by reducing the constraints within the assembly spaces and reconstructing these spaces to give novel molecules as an outcome of alternate selection pressure.

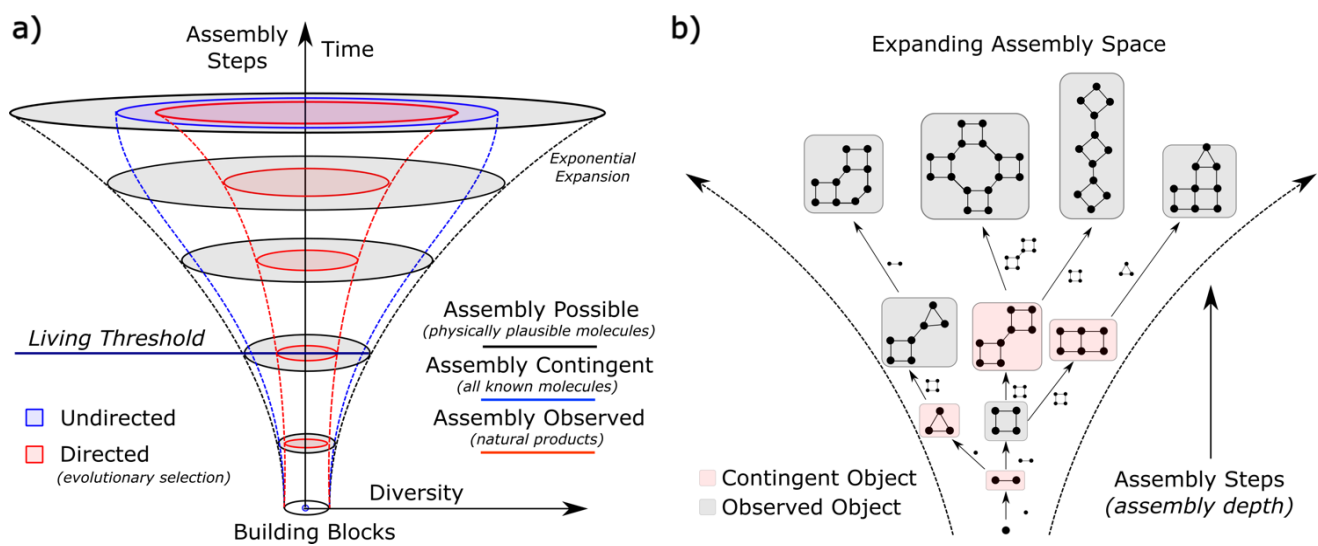

**Fig 1: Concept of assembly spaces.** (a) The figure shows the expansion of assembly spaces where Assembly Possible ($A_P$, shown in black) represents all physically plausible molecules, Assembly Contingent ($A_C$, shown in blue) represents all known molecules (such as PubChem database *(25)*), and Assembly Observed ($A_O$, Assembly Observed, shown in red) represents natural products (such as COCONUT database *(26)*). The living threshold represents the complexity threshold over which the observed molecule in high abundance signifies the presence of life. (b) Example of an expanding assembly space of an ensemble where at each assembly depth, observed and contingent objects exist. The assembly spaces $A_P$ and $A_C$ represent combinatorial spaces of physically plausible objects because of undirected and directed processes towards higher complexity such that $A_C$ is a subspace of $A_P$ ($A_C \subseteq A_P$). The assembly space $A_O$ ($A_O \subseteq A_C$) represents the space of observed objects (or experimentally measured) which are constructed as an outcome of selection and emerged in high copy number.

**Significance of Assembly Spaces**

The assembly space of natural products represents a set of complex and highly selective molecules whose physical construction pathways are too complex to be known as they include evolutionary processes over a very long timescale. At such complexity, the presence of the complex object itself at higher abundance represents the presence of evolutionary processes indicating life. In the context of assembly spaces *(13)*, the shortest construction pathway quantifies the minimal required information to build an object, where each step along the pathway $a \to a + 1$, i.e. where the assembly index goes up by one indicating the addition of a constraint. This approach captures the process of *selection* from the assembly pool to *combine* molecular substructures in a specific way which quantifies the presence of the evolutionary processes which drive biochemical systems. These construction pathways define the lower bound of the free energy required to construct the molecule with connectivity as the only known constraint and do not explicitly consider other physical constraints such as bond strength, environmental factors etc. The potential configurational space of *selection* and *combination* of objects is extremely large, and each step along this pathway adds constraints or contingency which captures the effect of external factors such as physical selection due to the influence of evolutionary dynamics. One can argue that in molecular assembly spaces with bonds as building blocks, there is a weak but finite selectivity where specific bonding constraints are favoured. By comparing the spaces, Assembly Possible for all molecules ($A_M$) and Assembly Contingent for natural products ($A_{NP}$), this weak selectivity can be decoupled from selectivity introduced by the evolutionary processes. Assuming a lack of knowledge of temporal information for complex molecule formation as well as any intermediate molecular substructures observed, each step along the assembly pathway quantifies information regarding *selection* and *combination* representing the effect of the presence of an evolutionary process in biology. These effects can be quantified more precisely if temporal information of intermediate species within the joint assembly space has been observed.

**Assembly Depth and Joint Assembly Spaces**

The assembly index quantifies the number of steps in a serial construction process, where all the sub-objects (also referred as contingent objects) are required to be constructed sequentially (13). However, the order of construction steps along the shortest path is not unique if two independent construction processes have been used. As an example, the string AABB can be constructed in a serial process by two possible pathways $A \rightarrow AA \rightarrow AAB \rightarrow AABB$ and $B \rightarrow BB \rightarrow ABB \rightarrow AABB$, where steps $A \rightarrow AA$ and $B \rightarrow BB$ can be perfomed independent of each other. Here, we define a quantity called assembly depth ($d$) which represents the shortest path when construction processes can occur concurrently ($A \rightarrow AA; B \rightarrow BB; AA + BB \rightarrow AABB$). In principle, the assembly depth of an object defines the lower bound on the assembly index ($a$) where forward processes are occurring. As an example, Fig. 2a shows an assembly pathway of a molecule (Adenine) as an object and bonds as building blocks. In the case of molecules, the assembly index ($a$) is also referred to as Molecular Assembly ($MA$). If the assembly pathway constitutes the addition of building blocks to one linearly growing object, the $MA$ is equivalent to the assembly depth ($d$). If instead, the assembly pathway connects two or more distinct objects which are not building blocks and can be constructed independently, the $MA$ will differ from the assembly depth. In cases where two or more objects were constructed on different branches, but by the same number of joining operations, their assembly depth will be identical. Thus, the introduction of assembly depth removes the need to assign an arbitrary order to the joining operations in an assembly pathway in cases where multiple equivalent solutions exist, by allowing concurrent processes. This is particularly useful for forward dynamics, where future combinatorial spaces need to be generated based on the existing information within the assembly spaces. Practically, the $MA$ is defined as the number of joining operations to construct a given molecule from its assembly pathway. The assembly depth in contrast is assigned recursively starting from the building blocks (assembly depth 0). Each subsequent object's assembly depth is calculated according to equation (2).

$$d = max(d_{f1}, d_{f2}) + 1 \qquad (2)$$

where $d_{f1}$ and $d_{f2}$ are the assembly depths of the two fragments that were connected to build a new object. Per definition, the assembly depth of building blocks is set to 0. In the case of multiple (not building block) fragments with the same assembly depth within an assembly pathway (existence of concurrent processes), the $MA$ is not equal to the assembly depth. For example, the $MA$ of Adenine is 7, but the Assembly Depth is 5, see Fig. 2a (see SI section 1 for details on the assembly pathway and assembly depth calculations).

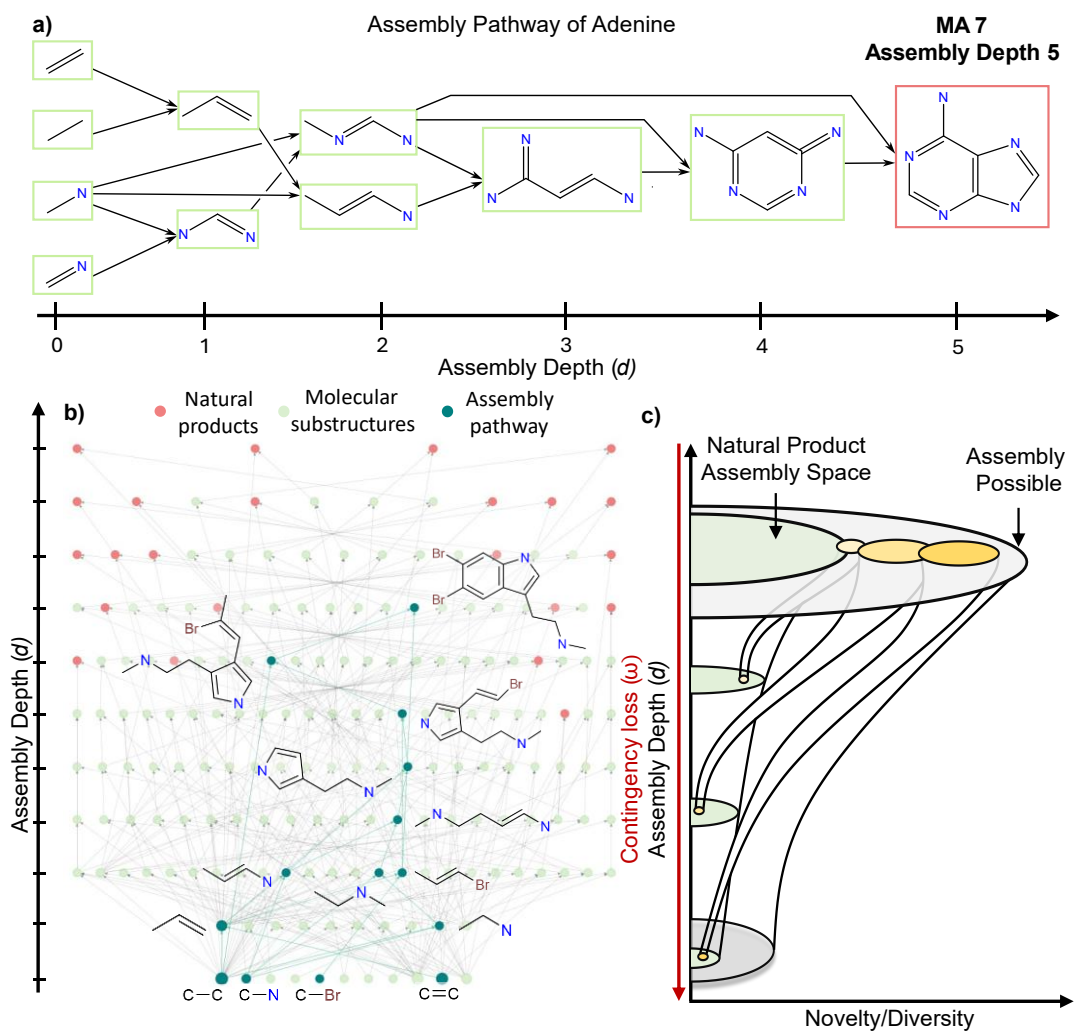

**Fig 2: Assembly Space of a single molecule and Joint Assembly Space of an ensemble of molecules. a)** The assembly pathway of adenine ($MA = 7, d = 5$). The unique bonds found in adenine act as the building blocks and are used together with the molecular substructures (both in green boxes) to construct adenine (red box). **b)** Depiction of the Joint Assembly Space (JAS) of 24 molecules from the COCONUT database (26) with a molecular assembly index between 7 and 10. The assembly fragments are ordered by their assembly depth. Each fragment is represented by a node in the JAS, where the size of the node represents the number of times the fragment was used during the construction process of the Assembly Space. Nodes that are coloured in light green represent building blocks and contingent objects (here molecular substructures) that have been used to construct the observed objects (natural products; red nodes). One observed molecules pathway is shown in dark green with the molecular (sub-)structures shown next to each respective node. **c)** Conceptual depiction of the assembly space of molecules. The outermost cone depicts the outer bounds of the Assembly Possible space, which contains all theoretical possible molecules that can be constructed from an assembly process obeying the laws of physics. The x-axis shows the distribution of novelty/diversity of observed objects at a given assembly step relative to an arbitrary reference molecule. Within Assembly Possible lies the assembly space of natural products (in green) which are the products of evolutionary dynamics of living systems. The yellow cones depict assembly spaces diverging from the assembly space of natural products at different assembly steps.

In this study, we used the natural product database (COCONUT database (*13*)) to generate the JAS as a model representing the outcomes of evolutionary processes found across biochemistry, where the JAS of 24 natural product molecules selected from the COCONUT database is shown in Fig. 2b. The JAS of all natural products signifies the presence of molecular evolution defined by the physical and chemical constraints given on Earth and the biological constraints of the evolutionary processes that produced all life on Earth. With 407,270 unique natural products, the COCONUT database is one of the most comprehensive databases of molecules that are a direct result of the evolutionary processes on Earth, accurately representing the chemical distribution found in biological life on Earth.

The JAS of natural products represents the assembly space of physically plausible molecules that emerged as an outcome of selection by the biological processes. This represents construction and selection processes leading to the observation of natural products utilising a common pool of building blocks and molecular intermediates and captures the effect of contingency in the assembly space, where the shared pathways represent utilisation of common molecular fragments and building blocks to create distinct molecules. An intuitive example of a molecular fragment adding contingency between many assembly pathways of natural products is the phenyl ring, which once constructed could be recursively utilized to create complex aromatic molecules such as phenolic compounds.

**Loss of Contingency and Reconstruction of Assembly Spaces**

The presence of contingency in the JAS is observed by the utilisation of sub-objects or molecular fragments in a recursive way to construct the observed molecules and represents the information existing as sub-objects that is accessible to construct the molecules. Given a JAS of some observed molecules, the existing contingency can be reduced in a restricted way by removing objects and sub-objects up to a certain assembly depth which we define as contingency loss. The remaining constraints can be used to construct new molecules different from the initial JAS. This approach is particularly important in exploring the assembly spaces defined by Assembly Possible constrained by the remaining Assembly Contingent. Fig. 2c shows the pictorial representation of assembly spaces $A_M$ (Assembly of all Possible Molecules) and $A_{NP}$ (Assembly Contingent or Observed) together with emerging novel spaces by losing contingency (represented by $\omega$) and reconstructing assembly spaces with alternate selection pressure.

Here, alternate selection pressure means sampling contingent objects and constructing physically plausible molecules with different rules compared to natural products. The key hypothesis is the higher the loss of contingency within the assembly space, the more novel molecules sharing the limited contingency can be generated with varying selection pressure. To study the influence of contingency loss on the emergence of novelty, an algorithm for the construction of molecules from a set of molecular fragments (i.e., in this case, an assembly space) was developed, see Fig. 3. Given an arbitrary JAS or an isolated assembly pathway, molecules were reconstructed from the available fragments, by first removing the $\omega$ topmost levels of assembly depth (for reconstruction from JAS or assembly pathway only) as an outcome of the removal of historical contingency. From the truncated JAS (JAS$_\omega$), molecules are then reconstructed by selecting a random fragment from the topmost assembly depth of JAS$_\omega$, and iteratively sampling other available fragments (compare Fig. 3a or SI Section 4 and 7 for details) with valence rules as the only constraint. In principle, the number of construction steps ($n_{steps}$) for the reconstruction of molecules can be set to any integer value. In this work, $n_{steps}$ was either set to a fixed integer value (for reconstruction from a single assembly pathway) or sampled according to the distribution of number of construction steps as observed by the observed objects in a JAS. Alternatively to starting the construction of molecules from a predefined JAS, it may also start from a set of building blocks (*S*) to iteratively build up a new JAS only constrained by the initial building blocks (Fig. 3b). In either case, an optional three-step filtering pipeline was implemented to reject chemically valid, but physically implausible molecules generated (Fig. 3c). In the filtering pipeline, molecules are first checked against a set of forbidden substructures as defined in MOLGEN *(27)*.

Conformers of molecules that pass this stage were generated and subsequently geometry optimised. Molecules for which either no valid conformer could be generated, or the geometry optimisation did not converge for any of the generated conformers were discarded.

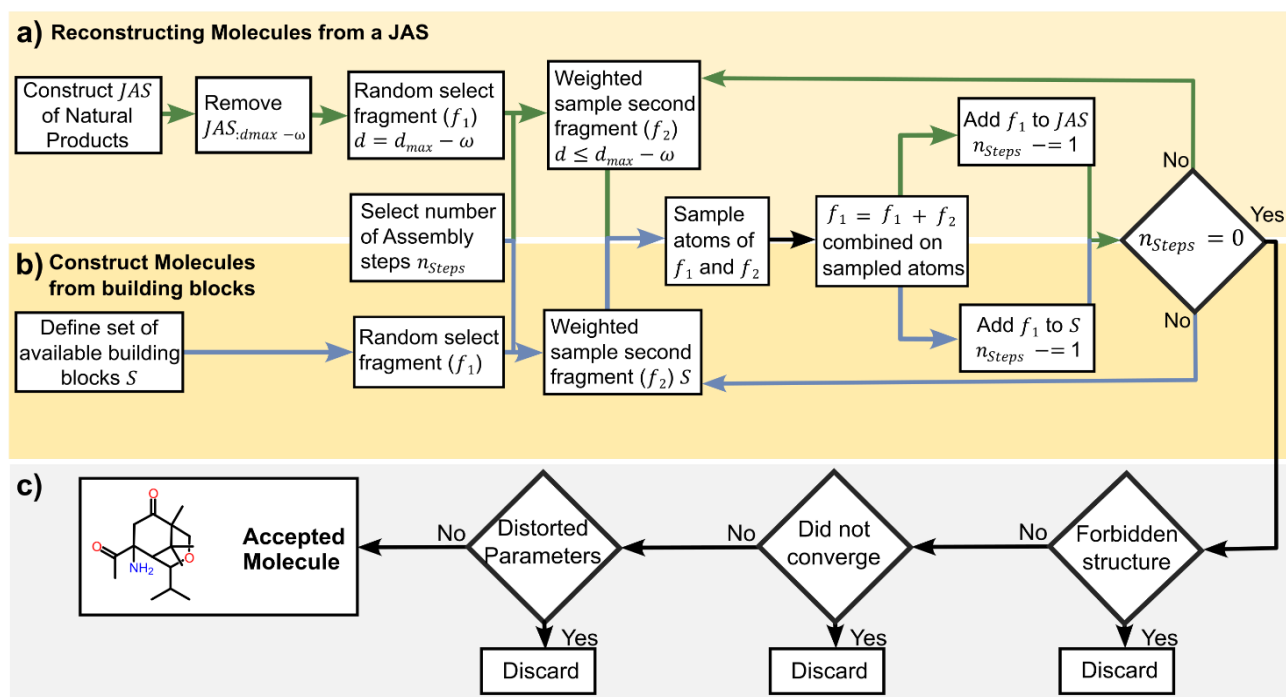

**Fig 3: Workflow for reconstructing and filtering molecules. a)** Reconstruction of molecules from a Joint Assembly Space. To reconstruct molecules (and a JAS) after contingency loss $\omega$, all fragments with an assembly depth ($d$) larger than $d_{max} - \omega$ are removed from the initial JAS. To start the reconstruction process, a fragment with assembly depth $d_{max} - \omega$ is randomly selected. The number of reconstruction steps is then sampled according to the distribution of construction steps in the original JAS. For each reconstruction step, a second fragment is sampled from the truncated JAS (JAS$_{dmax-\omega}$). To connect the two fragments, first, the number of connections to be formed is sampled. When two fragments have been successfully connected, the resulting object is added to the JAS. This process is repeated until $n_{steps}$ reconstruction steps have been completed. **b)** When constructing molecules from a set of predefined building blocks, first a building block is randomly chosen from the set of building blocks. The number of construction steps ($n_{steps}$) is defined by the user or application. For each construction step, a second building block from the set of building blocks is chosen. Fragments are connected as described for a). If the fragments were successfully connected, the new resulting fragment is added to the set of building blocks. The process is repeated until $n_{steps}$ have been completed **c)** To improve the physical validity of the resulting molecules an optional filtering pipeline was developed (see SI sections 4, 5 and 7 for details).

In the last stage, bond lengths and angles in passing molecules were compared against the distribution of the respective bond lengths and angles found in reported molecular structures from the CCDC database *(28)* (SI section 5 for details). The novelty of constructed molecules compared to the observed molecules in the original JAS or assembly pathway is then calculated by the Dice-Similarity between the Morgan-Fingerprints of two molecules (equation 3; SI section 6).

$$s = 2 \frac{|x_{f1} \cdot x_{f2}|}{|x_{f1}|^2 + |x_{f2}|^2} \tag{3}$$

where $s$ is the similarity between the Morgan Fingerprint between two molecules $f_1$ and $f_2$, $x_{f1}$ and $x_{f2}$ are the Morgan Fingerprints of the respective molecules. The novelty of molecule $f_1$ relative to molecule $f_2$ is defined as $n = 1 - s$.

To investigate the characteristic novelty or divergence of reconstructed molecules from the assembly pathway of a single molecule, 10000 molecules were reconstructed from the assembly pathway of Brefelamide (an aromatic amide isolated from Dictyostelium cellular slime molds see Fig. 4a) with increasing contingency loss $\omega$ from 1 to 10 (Fig. 4b). In good agreement with general intuition, the mean similarity between reconstructed molecules from the remaining object in the assembly pathway and Brefelamide decreases with an increase in contingency loss $\omega$ from approx. 0.8 ($\omega = 1$) to approx. 0.15 ($\omega = 10$). Notably, from contingency loss $\omega = 6$ to $\omega = 7$ the similarity of reconstructed molecules relative to Brefelamide is increasing. We hypothesize that this is an effect of phenol being *discovered* at assembly depth 3, which is the highest assembly depth after $\omega = 7$ and thus serves as the starting point for the reconstruction of new molecules. Since three phenol substructures are present in the Brefelamide the discovery, or lack thereof will strongly affect the similarity of resulting molecules. The similarity of reconstructed molecules from a given contingency loss relative to each other is shown as the mean maximal pairwise similarity in Fig. 4c. It can be observed that molecules, reconstructed from a shared assembly depth get more dissimilar to each other with increasing contingency loss i.e. increasing reconstruction pathway length.

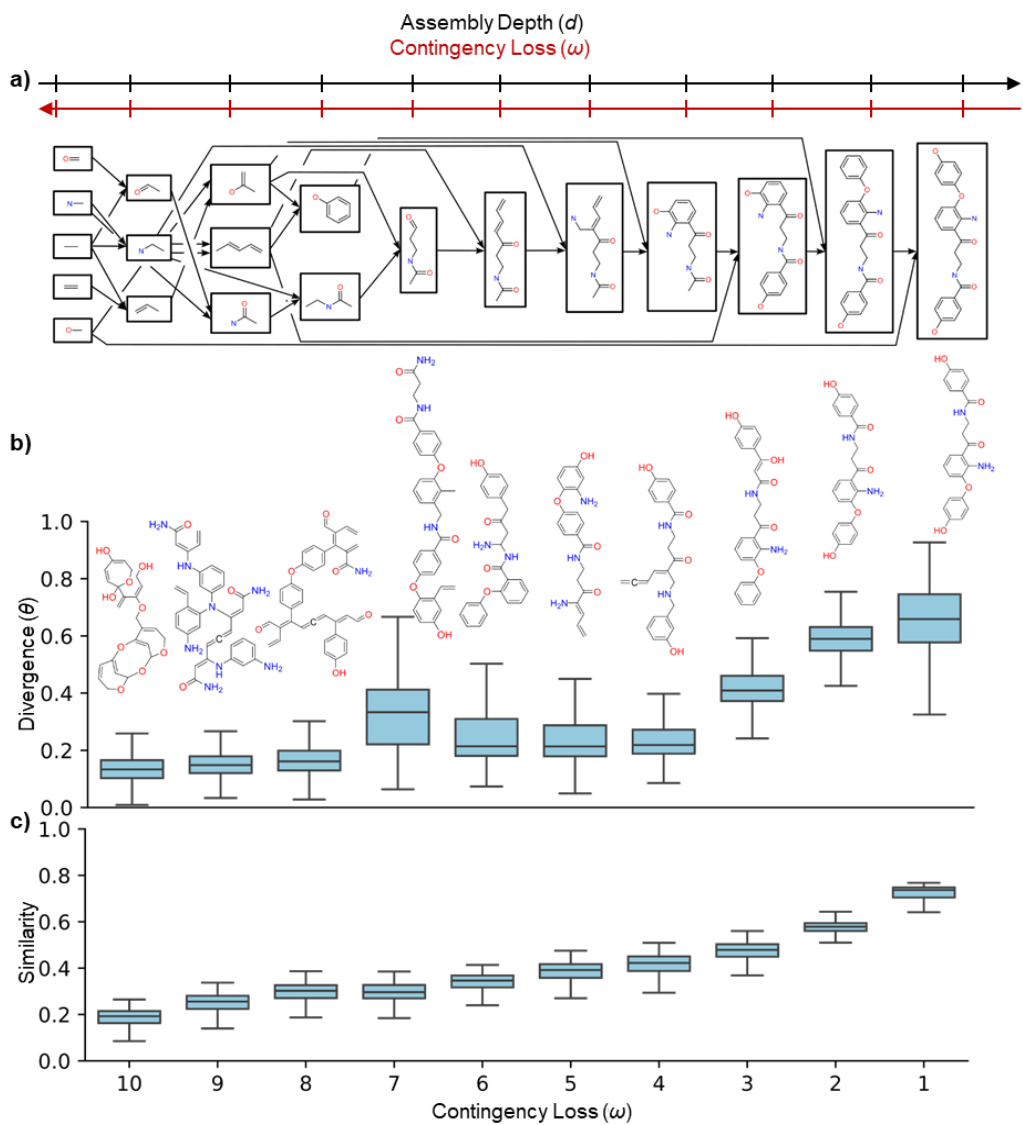

**Fig 4: Reconstruction of molecules from the assembly pathway of a single molecule. a)** The assembly pathway of the secondary metabolite Brefelamide (29) ordered by the assembly depth of the fragments within the pathway. **b)** Boxplot representation of the similarity of molecules reconstructed from the partial assembly path of Brefelamide, was calculated, by first removing $\omega = \{1, ..., 10\}$ contingency from the assembly path and reconstructing 10000 molecules from the remaining fragments up to the assembly depth of Brefelamide. The divergence of newly constructed molecules relative to Brefelamide was calculated as described in equation 3. The most similar reconstructed molecules after $\omega$ contingency loss steps are displayed after each respective bar plot. Error bars indicate the standard deviation of the divergence. **c)** Mean maximal pairwise similarity of the 10000 reconstructed molecules per $\omega$ relative to each other. The error bars represent the standard deviation of the mean maximal pairwise similarity.

**Generation of Novelty with Contingency Loss and Selection Pressure**

The generation of novelty in chemical spaces was studied based on the JAS of 211731 natural products from a COCONUT database (JAS$_{NP}$), with an assembly depth of up to 20 ($d_{max}$; SI section 2). To investigate the generation of novelty in newly generated JASs with respect to the parent JAS, an increasing amount of contingency ($\omega = \{2, 6, 10, 14\}$) was removed from JAS$_{NP}$ by removing objects (molecules and molecular fragments on their assembly pathways) with an assembly depth of $d_{max} - \omega$ as described in SI section 3.1. Starting from the resulting truncated JAS (JAS$_\omega$), the same number of molecules (observed objects) as removed through the contingency loss were reconstructed as described above (and SI section 4). The divergence of the reconstructed JAS (JAS$_{NP*}$) relative to JAS$_{NP}$ was calculated using equation 3 (compare Fig. 5; SI section 6 for more details). The divergence of two JASs is thus represented as the mean maximal similarity between molecules (and molecular substructures) of the same assembly depth and calculated for each assembly depth individually. While JAS$_{NP*}$ was reconstructed to mimic the construction process of the natural product JAS (JAS$_{NP}$) by setting the sampling weights in the molecule generation algorithm according to the distributions in JAS$_{NP}$ (SI Sections 3 and 7) an alternate selection pressure can be induced by tuning these sampling weights. In principle, the sampling weights can easily be adjusted to favour certain functional groups, atomic distributions, or any other chemical feature. Since only limited knowledge of the mechanistic leading to JAS$_{NP}$ is available the sampling weights for the fragment selection (SI section 3 and 7) were set to be uniform over all available fragments to model an alternate selection pressure. This represents a random exploration of the chemical space, contingent only on the previously explored fragments in JAS$_\omega$ (SI section 8.1). In both cases (reconstruction mimicking JAS$_{NP}$ and random reconstruction) the divergence of the resulting JAS (JAS$_{NP*}$) increases with increasing contingency loss $\omega$ relative to JAS$_{NP}$ (see Fig. 5). While for smaller contingency loss ($\omega = \{2, 6\}$) the divergence of the resulting JASs$_{NP*}$ relative to JAS$_{NP}$ is almost identical, an increasing difference of the divergence can be observed for $\omega = \{10, 14\}$ and increasing assembly depth when comparing random reconstruction vs reconstruction mimicking the dynamics in JAS$_{NP}$. Thus, the divergence of JASs, and the construction of molecules using fragments that exist in the evolutionary history of JASs, might be used to construct novel chemical spaces.

**Quantification of Selection in Earth's Chemical Space of Natural Products**

The exploration of an assembly space as an outcome of biological selection becomes highly restricted with increase in assembly depth due to informational constraints introduced by biological processes. Here, we quantify the overall selection by estimating how the exploration ratio of the assembly space of natural products with respect to all molecules scales with assembly depth by defining their relationship as $r = k\, e^{-\beta d}$, where $r$ is the exploration ratio, $k$ is the fixed ratio of the number of building blocks in natural products and all molecules, $\beta$ is the characteristic constant representing selection and $d$ is the assembly depth.

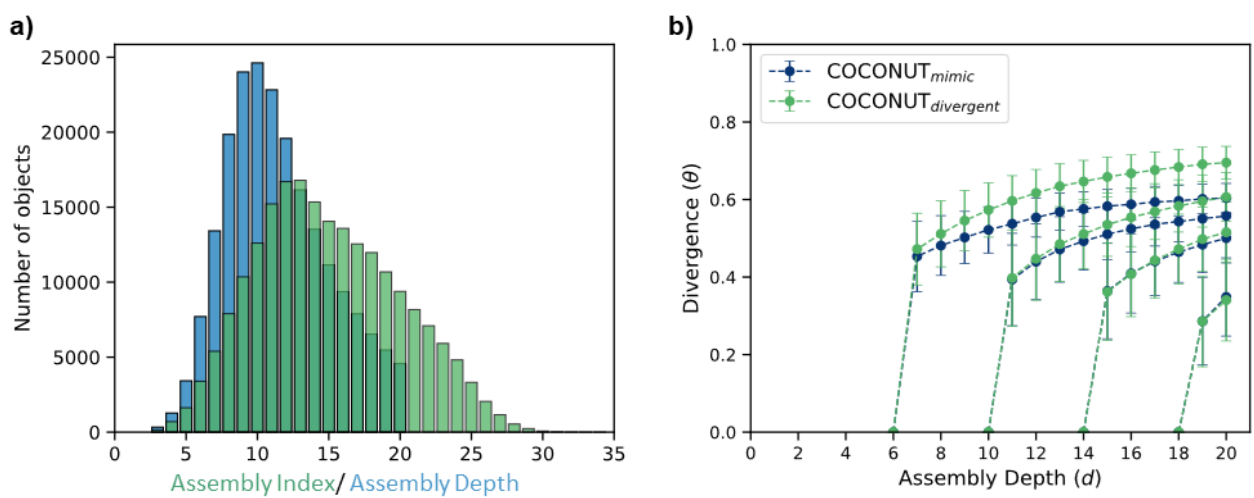

**Fig 5: Distribution of Assembly Depth and Assembly Index JAS$_{NP}$ and divergence of reconstructed JAS after contingency loss with alternate selection pressure. a)** Distribution of assembly index and assembly depth of the 211731 natural products with an assembly depth of up to 20 from which JAS$_{NP}$ was constructed. **b)** Divergence ($\theta$) of reconstructed JASs after contingency loss $\omega = \{2, 6, 10, 14\}$ relative to JAS$_{NP}$ calculated as described above. The JAS was either reconstructed to mimic JAS$_{NP}$ (COCONUT$_{mimic}$; blue) or diverge from JAS$_{NP}$ (COCONUT$_{divergent}$; green).

The molecular space was modelled using the JAS of more than 70 million molecules extracted from the PubChem database (JAS$_{PC}$) and the contingent assembly space as JAS of natural products (JAS$_{NP}$; see SI section 2 and 9 for more details on JAS$_{NP}$ and JAS$_{PC}$). JAS$_{NP}$ was constructed from all natural products with an assembly depth of up to 25 from COCONUT database. Here, JAS$_{NP}$ and JAS$_{PC}$ represent $A_{NP}$ and $A_M$ respectively where $A_{NP}$ is a subset of $A_M$ as described above. To quantify the selection that was required to construct $A_{NP}$ within $A_M$, all unique objects grouped by assembly depth were aggregated for both JAS$_{PC}$ and JAS$_{NP}$. The exploration ratio defined as $N_{NP}/N_{PC}$ of JAS$_{NP}$ relative to JAS$_{PC}$ was then calculated by taking the ratio of unique objects in JAS$_{NP}$ ($N_{NP}$) to JAS$_{PC}$ ($N_{PC}$) for objects of each assembly depth individually (see Figs. 6c and 6d). Additionally, the exploration ratio of reconstructed JAS (JAS$_{NP*}$) after contingency loss from JAS$_{NP}$ relative to JAS$_{PC}$ was calculated ($N_{NP*}/N_{PC}$). This was done by removing contingency from JAS$_{NP}$ as described above by either removing objects with an assembly depth larger than $d_{max} - \omega$ where $\omega = 23$ (see Fig. 6b and 6c), or removing the objects with an assembly depth larger than $d_{max} - \omega$ with $\omega = \{17, 19, 21\}$ as well as the objects exclusively appearing on their assembly pathways (see Fig. 6a and 6d). In this case, the ratio $k$ was adjusted to match the ratio of building blocks after contingency loss. Subsequently, molecules were reconstructed as described above (see SI section 4 for details). In the latter case, the reconstruction of molecules was controlled by adjusting the exponential scaling factor $s$ for the fragment selection $s = \{0.5, 1, 5\}$ as described in SI section 7. In short, adjusting this scaling factor results in a distribution shift where $s = 0$ represents a uniform distribution and increasing $s$, increases the likelihood of a fragment being sampled that was preferentially selected in the original JAS$_{NP}$.

As previously defined, the exploration ratio of JAS$_{NP}$ as well as all reconstructed JASs$_{NP*}$ was estimated with the exponential function $k\, e^{-\beta d}$ where $d$ is the assembly depth, the ratio $k \approx 0.12$, and $\beta$ is a characteristic constant quantifying the selection induced to attain a JAS subspace (see Figs. 6c and 6d and SI section 10). In case a JAS is fully explored $\beta$ will be 0. Thus, as the value of $\beta$ increases, the process must become increasingly selective in order to construct the corresponding JAS subspace. $\beta$ estimates the amount of selection required to construct the JAS of natural products (JAS$_{NP}$) within the JAS on a planetary scale (JAS$_{PC}$) to be 0.499.

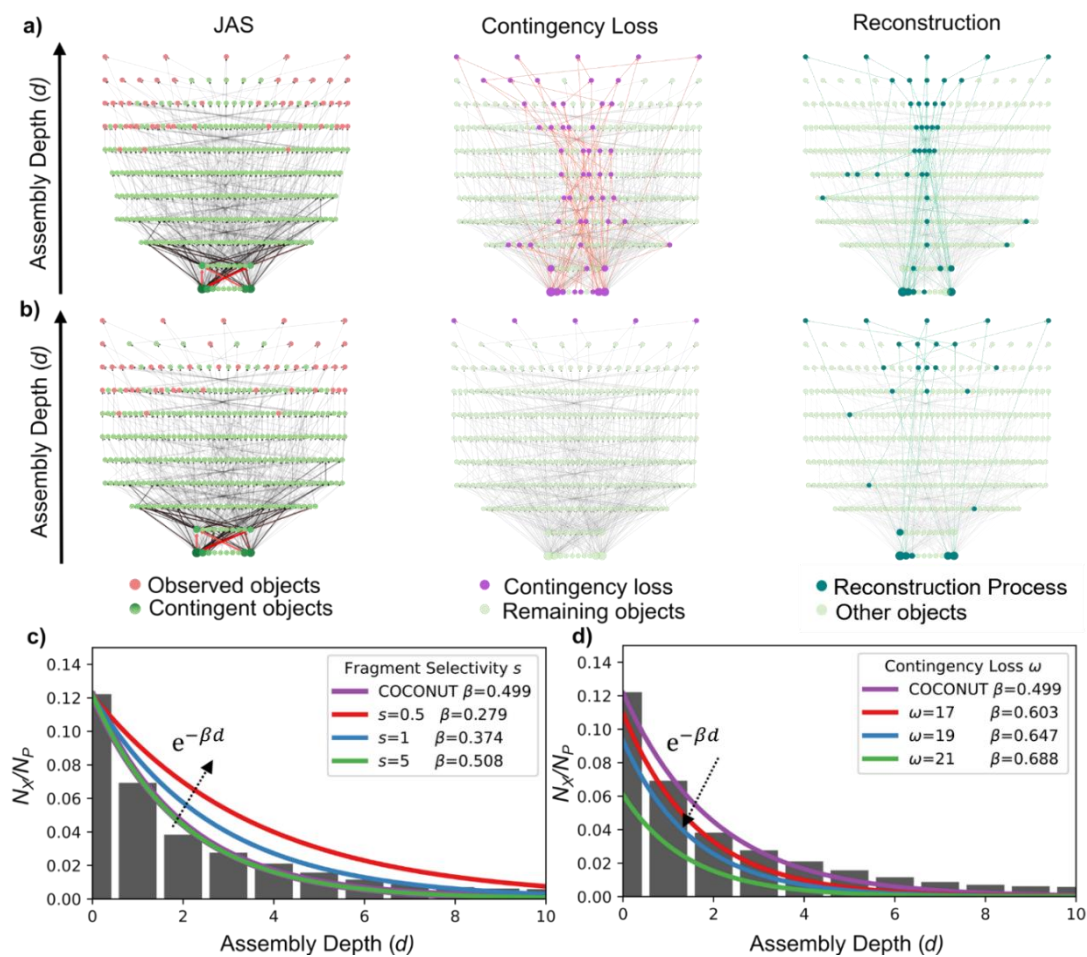

**Fig 6: Exploration of JAS$_{PC}$ by JAS$_{NP}$ and selectively reconstructed JASs. a)** Schematic representation of a JAS from which contingency is removed by removing observed molecules and the fragments on their pathways and subsequent reconstruction of the JAS (left to right). **b)** Schematic representation of JAS from which contingency is removed by removing objects with an assembly depth larger than $d_{max}$ and subsequent reconstruction of the JAS (left to right). **c)** Exploration ratio ($N_{NP}/N_{PC}$ or $N_{NP*}/N_{PC}$; the ratio of unique objects per assembly depth relative to JAS$_{PC}$) of JAS$_{PC}$ by the JAS of natural products (JAS$_{NP}$; grey bars) and reconstructed JASs thereof after contingency loss $\omega = 23$ by removing objects as described in b). The fragment selection for the reconstruction of new molecules was altered by adjusting the exponential scaling factor $s = \{0.5, 1, 5\}$ (see Equation 3 SI). The exploration ratio $N_X/N_P$ was estimated by an exponential function. The characteristic constant ($\beta$) of the exploration ratio by JAS$_{NP}$ was 0.499 (purple) and 0.279, 0.374, and 0.508 of the JAS with $s = \{0.5, 1, 5\}$ respectively (red, blue and green). **d)** $N_{NP}/N_{PC}$ of JAS$_{PC}$ by JAS$_{NP}$ (grey bars) and reconstructed JASs after contingency as described in a) with contingency loss $\omega = \{17, 19, 21\}$. The exploration ratio was estimated by an exponential function. The characteristic constant ($\beta$) of the exploration ratios was 0.603, 0.647, and 0.688 of the JAS with $\omega = \{17, 19, 21\}$ respectively (red, blue and green). The value of $\beta$ for JAS$_{NP}$ (purple) is equivalent to c) since these represent the same data.

The reconstruction of JASs from JAS$_{NP}$ with the same amount of contingency loss ($\omega = 23$) but differing exponential scaling factor for the fragment selection $s = \{0.5, 1, 5\}$ results in increasing exploration ratios, with decreasing $s$ ($\beta$ of 0.279, 0.374, and 0.508 respectively; Fig. 6c). Likewise, removing an increasing amount of contingency ($\omega = \{17, 19, 21\}$) with constant $s$ results in a decreasing exploration ratio with $\beta$ of 0.603, 0.647, and 0.688, respectively (Fig. 6d).

**Constructing a Drug-like Chemical Space from the Contingency of Natural Products**

In principle, every JAS that is the subset of another JAS can be constructed from this parent JAS if the exact dynamics and selection conditions are known. As an example, the JAS of natural products was constructed on Earth within Assembly Possible as defined by all known molecules. While we do not have exact knowledge of the dynamics and selection conditions leading to the natural products observed on Earth, the chemical space distribution of JAS of natural products can be reconstructed by mimicking the fragment utilization and construction step distribution observed during the reconstruction of molecules. Here, we postulate that this concept can be extended to construct drug-like molecules from the JAS of natural products since many drug molecules are derived from natural products which we consider to be a subspace of Assembly Contingent and Assembly Possible (see Fig. 7a). The exploration ratio of the JAS of all known molecules (JAS$_{PC}$) and JAS of natural products (JAS$_{NP}$) by the JAS of drug molecules (JAS$_D$) was calculated as described above (Fig. 7b). JAS$_D$ was modelled with 10656 small molecules from the ChEMBL database with an Assembly Depth of up to 20 (SI section 2). The exploration of JAS$_D$ was quantified by an exponential function ($e^{-\beta d}$), $\beta$ (0.358 and 0.849 for exploration of JAS$_{PC}$ and JAS$_{NP}$ respectively) but it does not quantify the selection induced to achieve the observed JAS$_D$ by an evolutionary process, however it quantifies the exploration of an assembly space through a selected subspace. Whereas we observe almost full exploration of JAS$_{NP}$ by JAS$_D$ for fragments of lower assembly depth, the exploration ratio rapidly decreases with increasing assembly depth. As expected, the exploration of JAS$_{PC}$ is substantially lower. As natural products naturally evolved to interact with biological organelles and biomolecules, fragments in the JAS of natural products thus may contain promising chemical substructures for novel drug-like compounds that are yet to be explored. By generating a JAS of drug-like molecules from the JAS of natural products, we thus may be able to generate molecules containing both classically drug-like features, but also chemical features produced by evolution and selection.

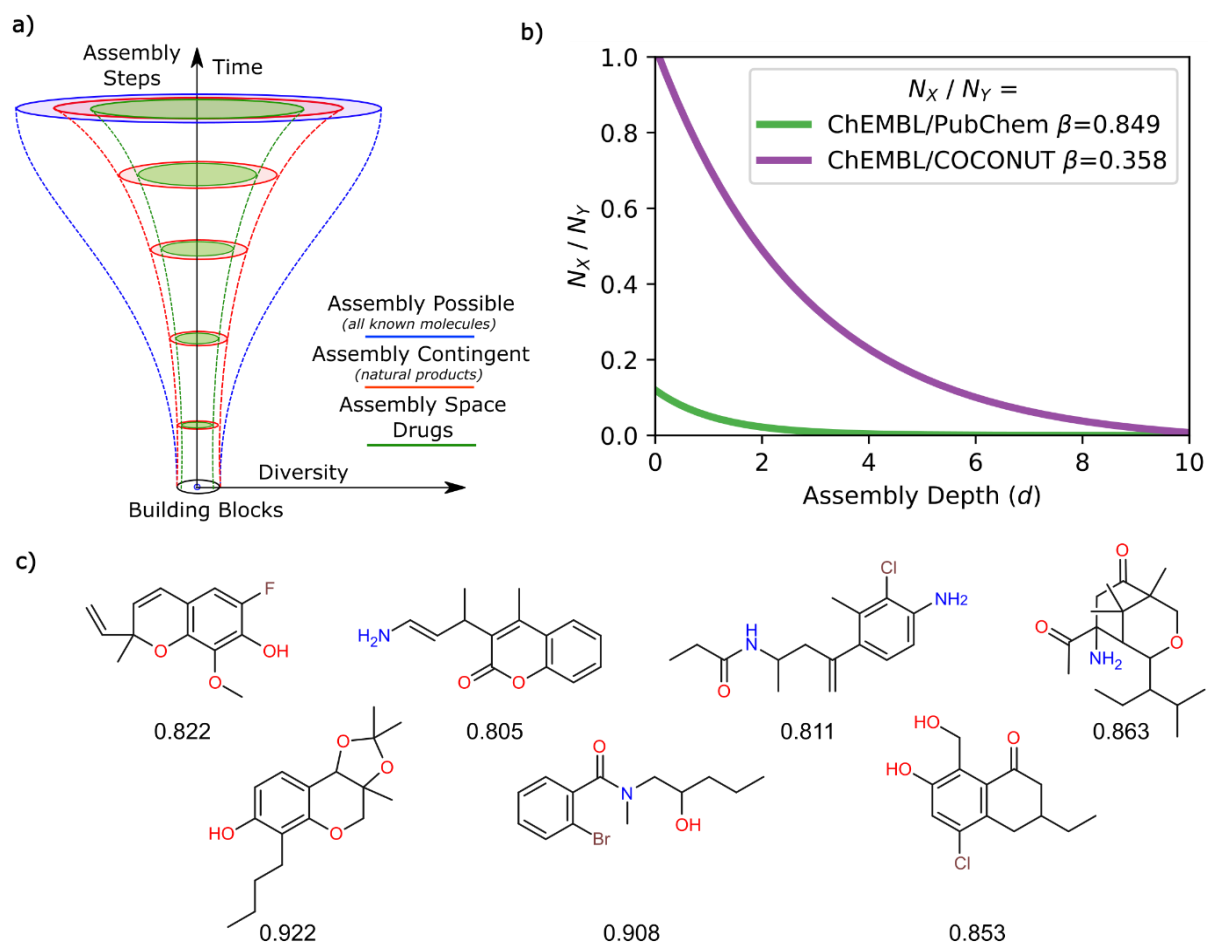

**Fig 7: Exploration of the Joint Assembly Space of drugs and construction of 7 drug-like molecules from natural product contingency. a)** Conceptual depiction of the JAS of drug molecules (JAS$_D$) in comparison to the JAS$_{NP}$ and JAS$_{PC}$. **b)** Exploration ratio ($N_X/N_Y$) from the fragment in JAS$_D$ (ChEMBL molecules) of JAS$_{NP}$ (COCONUT molecules), or JAS$_{PC}$ (PubChem molecules). The characteristic rate of $N_D/N_{NP}$ and $N_D/N_{PC}$ is 0.358 and 0.849, respectively. **c) 7** example molecules labelled with QED-score out of 10000 molecules that were reconstructed from the JAS of natural products after contingency loss $\omega = 15$ (down to assembly depth 5). In addition to the previously described substructure filters (see above), PAINS substructures (30) were used to filter out molecules with known problematic substructures in drug-screening assays.

To test this hypothesis, contingency was removed from the JAS of natural products by deleting molecular fragments and molecules with an assembly depth larger than 5. From the remaining fragments, 10000 molecules were reconstructed as described above (see SI section 8.2). Additionally to the filtering steps described previously, PAINS substructure filters (30) were used to filter out molecules commonly considered to be problematic in drug-screening campaigns. QED scores were calculated for the reconstructed molecules with 7 exemplary molecules shown in Fig. 7c showcasing a strategy for exploring novel drug-like molecules. Although the potential of drug-like molecules reconstructed from the JAS of natural products remains to be further explored and developed, this represents a potential workflow to explore chemical spaces guided by the constraints of evolution and selection that lead to physically observed molecules.

**Conclusions**

In this work, we have expanded the conceptual framework of AT *(13)* by introducing the significance of causal contingency in assembly spaces and its application to molecular spaces where the mechanisms of chemical reaction networks are too complex, occur over long timescales and are unknown. In the evolutionary systems, these contingent effects indicate the selection processes with the vast combinatorial universe which can be quantified. As an application of AT, we used natural products and all molecules' databases as observed and physically plausible molecular ensembles, generated their assembly pathways, and created joint assembly spaces which quantify the overall informational constraints required to construct them. By introducing the concept of contingency loss with the joint assembly spaces and developing a cheminformatic engine to filter physically plausible molecules, we demonstrated and quantified how novelty emerges in molecular space by losing causal contingency. Using this approach, we expanded and quantified the properties of assembly spaces of natural products creating physically plausible molecules in the presence of alternate selection pressure and at different degree of contingency loss. This approach is applicable to understanding the emergence of complex molecular observables in the presence of alternate reaction networks of the scale of biologically relevant evolutionary systems. We then compared the joint assembly spaces of natural products and all molecules' database to quantify the degree of selection emerging from the evolutionary processes on Earth and further expanded to explore drug-like molecules utilizing contingent space of natural products. This approach provides a novel framework to understand and quantify selection processes emerging from complex evolutionary networks and exploring alternate selection pressures and their observables beyond Earth biochemistry.

**Abbreviations**

The following abbreviations were used to describe the Joint Assembly Spaces explored in this work:

**JAS$_{PC}$** – *the JAS of all known molecules modelled by over 70 million molecules from the PubChem database*

**JAS$_{NP}$** – *the JAS of natural products as an outcome of the evolutionary machinery of Earth modelled by approx. 211 thousand molecules from the COCONUT database*

**JAS$_{NP*}$** – *JASs reconstructed from JAS$_{NP}$ after contingency loss*

**JAS$_D$** – *the JAS of drug molecules modelled by 10656 small molecules from the ChEMBL database*

**JAS$_\omega$** – *a JAS after loss of contingency (ω)*

Exploration ratios were abbreviated as follows:

$N_{NP}/N_{PC}$ – *the exploration ratio of JAS$_{PC}$ by the objects in JAS$_{NP}$*

$N_{NP*}/N_{PC}$ – *the exploration ratio of JAS$_{PC}$ by the objects in JAS$_{NP*}$*

$N_D/N_{PC}$ – *the exploration ratio of JAS$_{PC}$ by the objects in JAS$_D$*

$N_D/N_{NP}$ – *the exploration ratio of JAS$_{NP}$ by the objects in JAS$_D$*

## Methods

All the molecular assembly calculations on molecules from COCONUT and PubChem database were performed using AssemblyGo (12) and AssemblyCpp codes (which is a faster C++ implementation of AssemblyGo algorithm). All the pathways and further data analysis were performed using Python3 using standard libraries.

## Data Availability

All the codes used to perform analysis and generate figures in the manuscript and Supplementary Information are available at https://github.com/croningp/molecular_spaces.

## Author Contributions

L.C. conceived the idea and research plan together with A.S and S.P. A.S. built the concept and approach to utilise the methods of Assembly Theory for quantifying selection together with L.C. and S.P. S.P. and A.S. run the assembly calculations on the databases. S.P. developed cheminformatics tools and performed all the calculations on joint assembly spaces, molecular reconstruction and quantification of selection. A.S. and L.C. mentored S.P. All the authors wrote the manuscript.

## Acknowledgements

We acknowledge financial support from the John Templeton Foundation (grant nos. 61184 and 62231), the Engineering and Physical Sciences Research Council (EPSRC) (grant nos. EP/L023652/1, EP/R01308X/1, EP/S019472/1 and EP/P00153X/1), the Breakthrough Prize Foundation and NASA (Agnostic Biosignatures award no. 80NSSC18K1140), MINECO (project CTQ2017-87392-P) and the European Research Council (ERC) (project 670467 SMART-POM). We like to acknowledge Amit Kahana for the feedback on the manuscript.